\documentclass[aps,prb,amsmath,amssymb,reprint,showpacs]{revtex4-1}

\usepackage{dcolumn}
\usepackage{multirow}
\usepackage{graphicx}
\usepackage{bm}

\begin{document}
\title{Phases of the excitonic condensate in two-layer graphene}% Force line breaks with \\
\author{Yevhen F. \surname{Suprunenko}}
\email{y.suprunenko@lancaster.ac.uk}
\author{Vadim Cheianov}
\author{Vladimir I. Fal'ko}
\affiliation{Department of Physics, Lancaster University, Lancaster LA1 4YB, United Kingdom.}

\date{\today}

\begin{abstract}
Two graphene monolayers that are oppositely charged and placed
close to each other are considered. Taking into account valley and spin degeneracy of electrons we
analyze the symmetry of the excitonic insulator states in such a system and build a phase
diagram that takes into account the effect of the symmetry
breaking due to the external in-plane magnetic field and the carrier density imbalance between the layers.
\end{abstract}


\maketitle

\section{\label{sec:1 Introduction}Introduction.}
\begin{figure}[b]
\begin{center}
 \includegraphics[width=8cm]{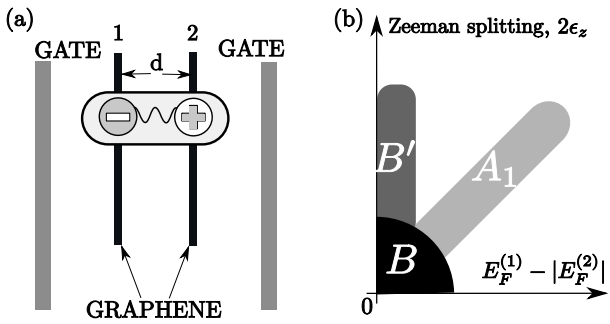}
\end{center}
\caption{
(a) The excitonic condensation due to an electron-hole
pairing is studied in the system of two spatially separated
graphene monolayers with an excess of electrons on layer 1 and a lack
of electrons on layer 2.
(b) The schematic phase diagram of the excitonic condensation
in the system at different values of a Zeeman splitting and different values of the asymmetry between
Fermi energies in layer 1 and 2, $\epsilon_{Z}=\mu_{B}|\mathbf{h}|$ is the Zeeman energy in an in-plane magnetic field $\mathbf{h}.$
} \label{fig 1}
\end{figure}

The excitonic insulator \cite{blatt,keldysh kopaev,jerome rice kohn
PR,keldysh kozlov,halperin rice rev mod phys} was predicted
theoretically four decades ago in 3D semiconductors and then in
spatially separated layers of electrons and holes.\cite{lozovik
yudson,shevchenko} Since then, an excitonic insulator has been
searched for in a variety of systems. The excitonic insulator is a material where the electron-hole excitonic correlations
lead to the formation of a gapped state characterized by the order
parameter resembling a superfluid condensate of excitons. Such a
correlated state has been observed in double-quantum well
semiconductor structures in quantizing magnetic fields.\cite{quantum well start,butov,moon,zhang joglekar
9,spielman,eisenstein,high} After the experimental
discovery of graphene \cite{Novoselov
1,Novoselov 2,Zhang 1,Zhang 2}
it has been discussed as a possible
candidate for experimental realization of the excitonic insulator state \cite{Aleiner,lozovik
sokolik,lozovik merkulova sokolik,min .. macdonald,zhang joglekar,lozovik willander}
sparking the on-going debate \cite{kharitonov efetov,KharitonovEfetov0903,macdonald
comment} about the critical temperature $T_{c}$ of the excitonic
condensate transition in a two-layer graphene system.
Various estimations of $T_{c}$ for such a system lay in a wide region of magnitudes from milli-Kelvins\cite{kharitonov efetov,KharitonovEfetov0903} up to Kelvins\cite{lozovik2009,lozovik2009v2,lozovik2010,mink} and further up to the room temperature.\cite{min .. macdonald,zhang joglekar}
The considered system \cite{lozovik
sokolik,lozovik merkulova sokolik,min .. macdonald,zhang
joglekar,kharitonov efetov,macdonald comment,KharitonovEfetov0903} consists of two parallel, separately controlled graphene
monolayers, in which
external gates induce a
finite density of electrons in the layer 1 and holes in the layer
2, Fig. \ref{fig 1}(a).
Recently the two-layer graphene system has been obtained
experimentally.\cite{schmidt,schmidt2,schmidt3,schmidt4,FalkoGaugeField,FalkoCheianovTunable}

In this paper, we extend the existing theory of the excitonic
insulator state in a two-layer graphene system:
we analyze a symmetry of the excitonic insulator and classify its phases.
As a result a phase diagram of the excitonic insulator is built, that takes into account the effect of the
symmetry breaking due to the Zeeman splitting and the asymmetry
between electron/hole densities in the layer 1 and 2. A phase
diagram, Fig. \ref{fig 1}(b), contains 3 phases: ${\rm B,B^{\prime}}$ and ${\rm A_{1}}.$
Transitions between phases are found to be of the first order.
These transitions are subject to the use of an in-plane magnetic
field and a variation of external gate voltages, leading to
different charge carriers densities in layers: the density of all
electrons $n_{1e}$ in layer 1 (which corresponds to the Fermi
energy $E_{F}^{(1)}=\hbar v\sqrt{\pi n_{1e}}/2),$ and density of holes
$n_{2h}$ in layer 2 (which corresponds to the negative Fermi
energy in the layer 2, $E_{F}^{(2)}=-\hbar v\sqrt{\pi n_{2h}}/2$).

The ${\rm B}$ phase, Fig. \ref{fig 1}(b), exists when there is no magnetic field and charge carriers densities are the same in both layers $n_{1e}=n_{2h}$ (i.e. when $E_{F}^{(1)}=|E_{F}^{(2)}|).$
The ${\rm B^{\prime}}$ phase exists at the same condition $n_{1e}=n_{2h}$ but when an in-plane magnetic field is applied, which causes a Zeeman splitting of energies of electrons with different spin projections.
The ${\rm A_{1}}$ phase exists when a symmetry of charge carriers density is violated, e.g. $n_{1e}>n_{2h},$ and when the corresponding splitting of Fermi energies $E_{F}^{(1)}-|E_{F}^{(2)}|$ is equal to the Zeeman splitting due to an in-plane magnetic field.

The diversity of obtained phases is due to the high symmetry
of the normal ground state, which can be broken in several
different ways leading to a variety of phases possessing
different symmetry groups. A well known example of the system with the diversity of
phases due to various normal state symmetry breaking is liquid
Helium-3.\cite{leggettrmp75,wheatley,mineevufn,voloviksymmetryin3-Hechapter,vollhardt wolfle}
In liquid Helium-3 the symmetry of the order
parameter can be changed by correspondent external parameters,
leading to phase transitions.

The analysis in this paper is organised as follows. Section \ref{sec:2 Two-layer Hamiltonian}
describes the theoretical model of the considered system.
Pairing of electrons and holes within mean field theory is introduced in the Section \ref{sec:3 Excitonic pairing}. Section \ref{sec:4 Symmetry} provides symmetry analysis and the phase classification of the excitonic correlated state. Section \ref{sec:5 Phases} contains detailed description of the most symmetric phases, their properties are summarizes in Table \ref{tab:phases}. Results are discussed in Section \ref{sec:6 Results}.

\section{\label{sec:2 Two-layer Hamiltonian}Two-layer Hamiltonian}

\begin{figure}[h!]
\begin{center}
 \includegraphics[width=5cm]{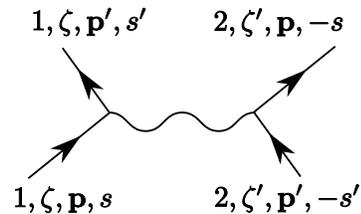}
\end{center}
\caption{ A typical transition which is described by
$\hat{H}_{12}$ in Eq. (\ref{H int}). Indices
$l=1{\rm~or~}2,\zeta,\mathbf{p},s$ denote a layer, a pair of the spin projection
and valley, a momentum of an electron and a conduction $s=+$  or valence $s=-$
band. }
 \label{fig:Vrpa}
\end{figure}

Graphene \cite{wallace,review 1,review 2} is a gapless
semiconductor with the Fermi surface consisting of two distinct
points, $\mathbf{K}_{+}$ and $\mathbf{K}_{-},$ called valleys.
Near these Fermi points electrons have a linear dispersion $E(p)=\pm
vp,$ with a velocity $v\approx10^{8}{\rm cm/sec},$\cite{Novoselov
2} here $p=|\mathbf{p}|,$ $\mathbf{p} =\mathbf{k}-\mathbf{K}_{\pm}$ is the momentum of an electron relative to the Fermi point.
Using external gates, one can independently tune the carrier
density in each of the two graphene flakes.\cite{Novoselov 2}
Neglecting tunneling, the electrons in the two layer graphene
system initially can be described with the Hamiltonian
$\hat{H}_{{\rm 2layer}}=\hat{H}_{{\rm s.p.}}+\hat{H}_{11}+\hat{H}_{22}+\hat{H}_{12},$
here the single particle part of the Hamiltonian is

\begin{equation}
 \hat{H}_{{\rm s.p.}}=
 \sum_{l,\zeta,\mathbf{p},s}
   (s vp-E_{F}^{(l)})\,\,
   a^{\dagger}_{l,\zeta,\mathbf{p},s}\,\,
   a_{l,\zeta,\mathbf{p},s},
   \label{H kin}
\end{equation}
the operators
$a_{l,\zeta,\mathbf{p},s}^{\dagger}\,(a_{l,\zeta,\mathbf{p},s})$
create (annihilate) an electron on the $l=1,2$ layer on the
$s=+/-$ conduction or valence band with momentum
$\mathbf{p}=p(\cos\phi_{\mathbf{p}},\sin\phi_{\mathbf{p}}),$
$E_{F}^{(1)}$ and $E_{F}^{(2)}$ are the Fermi energies, which correspond to
charge carrier densities in the layers. The index $\zeta$ denotes
4 different pairs of spin projection $(\uparrow,\downarrow)$ and
valleys $(\mathbf{K}_{+},\mathbf{K}_{-}).$ In the Hamiltonian
$\hat{H}_{{\rm 2layer}}$ the terms $\hat{H}_{11}$ and $\hat{H}_{22}$ take into
account the intra-layer interaction. These terms can be ignored in
the following studies, provided that one uses a screened inter-layer
interaction in the term $\hat{H}_{12}.$ Hence in $\hat{H}_{12}$ we keep
only those terms that contribute to the BCS mean field theory,\cite{mineevsamokhin}
absorbing other contributions into a renormalization of the velocity
and the Fermi energy in the single particle part (\ref{H kin}) of the Hamiltonian

\begin{eqnarray}
 \hat{H}_{12}&=&-
  \sum_{\mathbf{p},\mathbf{p}^{\prime},\,s,s^{\prime}}
   V(|\mathbf{p}-\mathbf{p}^{\prime}|)\,
   {1+ss^{\prime}
   \cos(\phi_{\mathbf{p}}-\phi_{\mathbf{p^{\prime}}})
   \over2}
   \nonumber
\\
   &&\times
 \sum_{\zeta,\zeta^{\prime}}\,
   a^{\dagger}_{1,\zeta,\mathbf{p},s}\,\,
   a^{\dagger}_{2,\zeta^{\prime},\mathbf{p^{\prime}},-s^{\prime}}\,\,
   a_{1,\zeta,\mathbf{p^{\prime}},s^{\prime}}\,\,
   a_{2,\zeta^{\prime},\mathbf{p},-s}.
   \label{H int}
\end{eqnarray}
The scattering process, described by $\hat{H}_{12},$ is shown on
Fig. \ref{fig:Vrpa}.
The function $V(q)$
denotes a screened Coulomb interaction in
the static limit $V(q)=V(q,\omega\ll q)$. The factor $\left[1+ss^{\prime}
\cos(\phi_{\mathbf{p}}-\phi_{\mathbf{p^{\prime}}})\right] /2$
 in Eq.(\ref{H int}) reflects chiral properties of electrons related to the sublattice
composition of electronic Bloch wave functions.\cite{review 1,review 2}
These chiral properties of electrons result in the suppressed backwards scattering if
an electron does not change the energy band upon scattering $(ss^{\prime}=+),$ otherwise $(ss^{\prime}=-)$ the electron can not
forward-scatter.\cite{katsnelson}

\section{\label{sec:3 Excitonic pairing}Excitonic pairing, mean field order parameter}

The excitonic insulator state of the electron-hole liquid is
characterized by the electron-hole correlations on the Fermi
surface, Fig. 3. Mathematically it means that in the excitonic insulating
state there is a non-vanishing ground state average ${\rm F}$ of electron operators

\begin{figure}[b]
\begin{center}
 \includegraphics[width=8.0cm]{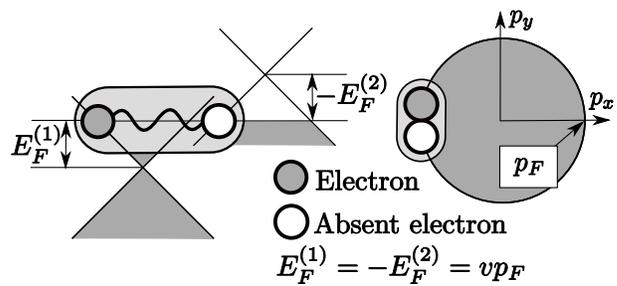}
\end{center}
\caption{
The excitonic electron-hole bound state in the two-layer graphene.
The left hand side of the figure shows the electron's spectrum in graphene layer 1 and 2. An electron on the Fermi surface in the layer 1 is shown as a fulfilled circle. Absence of an electron on the Fermi surface in layer 2 is shown as an empty circle. Closed line around both circles represents an excitonic pairing, which is developed due to a Coulomb interaction (shown as a wavy line).
The right hand side of the figure shows the coincided Fermi circles in both layers at Fermi momentum $p_F.$
}
\label{fig:phase}
\end{figure}

\begin{equation}
 {\rm F}_{\zeta\zeta^{\prime},s}({\bf p})
 =
 \langle
 {a}^{\dagger}_{2,\zeta^{\prime},{\bf p},-s}
 {a}_{1,\zeta,{\bf p},+s}
 \rangle.
 \label{anom average}
\end{equation}

For the existence of the non-zero ground state average ${\rm F}$ it is crucial that Fermi surfaces for electrons and holes coincide.\cite{BCS,mineevsamokhin}
Due to the electron-hole symmetry of the energy spectrum in graphene, the electron-hole excitonic correlations (\ref{anom average}) in the considered system are most developed when
the density of electrons in the layer 1 is equal to the density of holes
in the layer 2, $n_{1e}=n_{2h},$ or, in terms of Fermi energies $E_{F}^{(1)}=-E_{F}^{(2)}$, Fig.\ref{fig:phase}.
However apart from this condition there can be  certain other external conditions when excitonic correlations (\ref{anom average}) can be developed.
Thus, although the excitonic insulator state disappears when the symmetry $n_{1e}=n_{2h}$ is violated by external gates,
we show below that excitonic correlations can be restored by the in-plane magnetic field.
Based on the detailed analysis of excitonic correlations in monolayer graphene in the in-plane
magnetic field, which is done by Aleiner and co-authors\cite{Aleiner}, we show that the excitonic insulator state can exist in various phases in the two-layer graphene system.

In order to study phases of the excitonic insulator state at different external conditions, firstly we apply the standard mean-field approximation.\cite{mineevsamokhin} We assume that the product of the operators
$a^{\dagger}_{2,\zeta^{\prime},\mathbf{p},-s}a_{1,\zeta,\mathbf{p},s}$
weakly deviates from its non-vanishing ground state average. We expand
the interacting part $\hat{H}_{12},$ Eq. (\ref{H int}), of the Hamiltonian $\hat{H}_{{\rm 2layer}}$ up to the linear order with respect to these small deviations and neglect constant terms. The mean field Hamiltonian of the
system becomes

\begin{equation}
 \hat{H}_{{\rm mf}}=
  \hat{H}_{{\rm s.p.}}
  +
  \sum_{\mathbf{p},s,\zeta,\zeta^{\prime}}
  [
    a^{\dagger}_{1,\zeta,\mathbf{p},s}
    \Delta_{\zeta\zeta^{\prime},s}(\mathbf{p})
    a_{2,\zeta^{\prime},\mathbf{p},-s}
  +{\rm H.c.}],
  \label{mean-field H}
\end{equation}
where ${\rm H.c.}$ stands for "Hermitian conjugate", and

\begin{eqnarray}
 \Delta_{\zeta\zeta^{\prime},s}(\bf{p})&=&
  -\sum_{\mathbf{p^{\prime}},s^{\prime}}
{\rm F}_{\zeta\zeta^{\prime},s^{\prime}}({\bf p}^{\prime})
   V(|\mathbf{p}-\mathbf{p^{\prime}}|)\,
 \nonumber
 \\
 & &\times  {1+ss^{\prime}\cos(\phi_{\mathbf{p}}-\phi_{\mathbf{p^{\prime}}})
   \over2}.
  \label{Delta difinition}
\end{eqnarray}
Quantities $\Delta_{\zeta\zeta^{\prime},s}(\mathbf{p})$ form the matrix ${\Delta}$ of the
order parameter. Index $\zeta$ denotes 4 different pairs of spin projections and
valleys $(\uparrow \mathbf{K}_{+},\uparrow \mathbf{K}_{-},\downarrow
\mathbf{K}_{+},\downarrow \mathbf{K}_{-}).$
Thus in the spin$\otimes$valley space the order parameter is given by the
$4\times4$ matrix $\Delta$ with matrix elements given by (\ref{Delta difinition}).
For brevity we omit index $s$ and momentum $\mathbf{p}$ in the notation for the order parameter $\Delta.$

For further analysis it is convenient to rewrite the Hamiltonian (\ref{mean-field H}) as follows:
$\hat{H}_{{\rm mf}}=\sum_{\zeta,\bf{p},s}
 {\Psi}_{\zeta,\bf{p},s}^{\dagger}
 H_{{\rm mf}}(\mathbf{p},s)
 {\Psi}_{\zeta,\bf{p},s},$
where
$\Psi_{\zeta,\bf{p},s}=\left({a}_{1,\zeta,\bf{p},+s},{a}_{2,\zeta,\bf{p},-s}\right)^{T},$
and

\begin{equation}
 H_{{\rm mf}}(\mathbf{p},s)
 =
 \left(\begin{array}{cc}
 (svp-E_{F})\openone &\Delta\\
 \Delta^{\dagger}&-(svp-E_{F})\openone\\
 \end{array}\right).
 \label{Hmf 1/N}
\end{equation}
Here all elements of the matrix $H_{{\rm mf}}$ are $4\times4$ matrices in the spin$\otimes$valley space:
diagonal elements have structure of the identity matrix $\openone$ in this space, whereas $\Delta$ is given by some $4\times4$ matrix, whose structure is identified in this paper for each phase of the excitonic correlated state.
The matrix of the order parameter ${\Delta}$ describes the correlations between
conduction/valence electrons in the layers 1 and 2 below a critical
temperature $T_{c}$.

Nevertheless, the phase classification can be made regardless of the value of the transition temperature $T_{c}.$
Assuming that the excitonic insulator state can be observed in two-layer graphene system, we analyze the symmetry of the mean field Hamiltonian (\ref{mean-field H}) and the order parameter $\Delta.$ As a result the classification of all phases of the excitonic insulating state of the two-layer graphene system is presented in the next Section, and a detailed discussion of each phase is presented in Section \ref{sec:5 Phases}.

\section{\label{sec:4 Symmetry}Symmetry analysis of the correlated state}
The analysis in this section is based on the idea of
breaking of the initial symmetry of the hamiltonian by
the order parameter. The initial symmetry group $G$ of the Hamiltonian $\hat{H}_{{\rm 2layer}}$ is formed by global unitary transformations of an electronic
single-particle state in the 4-component spin$\otimes$valley space independently
in the layer 1 and 2. These transformations are represented by
independent matrices ${\rm U}^{(1)}$ and ${\rm U}^{(2)}$ in
layer 1 and 2 respectively. Therefore the group $G$ is given by
the direct product of corresponding unitary groups $U_{4}$

\begin{equation}
 G=
 U_{4}^{(1)}
 \times
 U_{4}^{(2)}.
 \label{G}
\end{equation}
Unitary group $U_{4}^{(l)},$ $l=1,2,$ consists of $4\times4$ unitary matrices ${\rm U}^{(l)}$
which perform transformations of electron's operators in the $l-$th layer as
follows:

\begin{equation}
    a_{l,\zeta,\mathbf{p},s}
    \rightarrow
    \sum_{\zeta^{\prime}}
    {\rm U}^{(l)}_{\zeta\zeta^{\prime}}
    a_{l,\zeta^{\prime},\mathbf{p},s}.
 \label{G transformations}
\end{equation}
Thus, as it is seen from Eqs. (\ref{mean-field H}) and (\ref{Hmf 1/N}), under symmetry transformations (\ref{G transformations})
the order parameter ${\rm \Delta}$ transforms as:

\begin{equation}
 {\Delta}
 \longrightarrow
 {\rm U}^{(1)\dagger}
 \,
 {\Delta}
 \,
 {\rm U}^{(2)}.
 \label{cond on Delta1}
\end{equation}
This implies that the Hamiltonian of the system is not invariant under the action of the group $G$ any longer.
However for any fixed non-zero ${\Delta}$
there is always some subgroup $H$ of the group $G,$ $H\subset G,H\neq G,$ such that all transformations from the group $H$ do not transform $\Delta,$ i.e. the order parameter $\Delta$ remains invariant:

\begin{equation}
 {\rm U}^{(1)\dagger}_{H}
 \,
 {\Delta}
 \,
 {\rm U}^{(2)}_{H}
 =
 \Delta.
 \label{cond on Delta U1U2H}
\end{equation}
Such transformations ${\rm U}^{(1)}_{H}$ in layer 1 and ${\rm U}^{(2)}_{H}$ in layer 2 form a symmetry group $H$
\begin{equation}
    \left(\begin{array}{cc}
    {\rm U}^{(1)}_{H}&{\rm 0}\\
    {\rm 0}&{\rm U}^{(2)}_{H}\\
    \end{array}\right)
 \in
 H\subset G.
 \label{U1U2 H}
\end{equation}
Only transformations from the group $H$ leave the ground state of the
excitonic insulator invariant, i.e. only these transformations leave the mean field Hamiltonian (\ref{mean-field H}) and (\ref{Hmf 1/N}) invariant:
\begin{equation}
    \left(\begin{array}{cc}
    {\rm U}^{(1)\dagger}_{H}&{\rm 0}\\
    {\rm 0}&{\rm U}^{(2)\dagger}_{H}\\
    \end{array}\right)
H_{{\rm mf}}(\mathbf{p},s)
    \left(\begin{array}{cc}
    {\rm U}^{(1)}_{H}&{\rm 0}\\
    {\rm 0}&{\rm U}^{(2)}_{H}\\
    \end{array}\right)
=H_{{\rm mf}}(\mathbf{p},s).
\label{Hmf H sym}
\end{equation}
Thus the symmetry group $G$ of the initial uncorrelated normal ground state of the system is broken down to the symmetry group $H$ of the ground state of the excitonic insulator.

All transformations from $G,$ which are not included in $H,$
form the factor-space $G/H.$
These transformations change the order parameter $\Delta$, however they do not change the energy of the corresponding ground state.
Therefore the manifold of all matrices $\Delta,$ which can be obtained by transformations from $G/H,$ form a degeneracy space of the order parameter. Consequently the manifold of the correspondent ground states form a phase of the correlated state.
It is important to notice, that all these ground states within the same phase are described by the same symmetry group $H,$ which is a symmetry group of the order parameter. Therefore phases of a correlated state can be classified by the symmetry group $H$ and the degeneracy space of the order parameter.

The phase classification presented in this paper is also reminiscent of the
classification of the various degeneracy spaces of the order
parameter in liquid Helium-3.\cite{leggettrmp75,wheatley,vollhardt wolfle,mineevufn} This
classification principle was used in the determination of
superconducting phases in nontrivial superconductors \cite{mineevsamokhin} and superfluid phases in liquid Helium-3.
\cite{brudervollhardt,voloviksymmetryin3-Hechapter}

\begin{table*}
 \caption{The phase classification of the
excitonic insulating state with respect to following two external
parameters:  (1) Zeeman energy $\epsilon_{Z}=\mu_{B}|\mathbf{h}|$
in an in-plane magnetic field $\mathbf{h},$ and (2) an asymmetry
between the electron density $n_{1e}$ in the layer 1
and the hole density $n_{2h}$ in the layer 2.
}
\begin{ruledtabular}
\begin{tabular}{c|c|c|c|c|c|c}
 \label{tab:phases}
 $\begin{array}{c}
        {\rm External}\\
        {\rm conditions}\\
  \end{array}$&
  $\begin{array}{c}
        {\rm Symmetry~group}~G\\
        {\rm of~the~two-layer}\\
        {\rm~Hamiltonian}
        \\
  \end{array}$
  &
  $~{\rm Phase}~$
  &
  $\begin{array}{c}
  {\rm Matrix~structure}\\
  {\rm of~the~order}\\
  {\rm parameter~}{\Delta}\\
        ({\rm here~~V},{\rm \widetilde{V}}\in U_{4})\\
   \end{array}$
  &
  $\begin{array}{c}\\
        {\rm Symmetry~group}~H\\
        {\rm of~the}\\
        {\rm order~parameter~}{\Delta} \\\\
  \end{array}$
  &
  $~{\rm dim}[G/H]~$
  &
  $\begin{array}{c}
        {\rm Single}\\
        {\rm particle}\\
        {\rm spectrum}\\
   \end{array}$\\
%%%%%%%%%%%%%%%%%%%%%%%%%%%%%%%%%%%%%%
\hline \hline
  &
  &
  $\begin{array}{c}\\  B\\ \\ \end{array}$
  &
  ${\rm V}$
  &
  $U_{4}^{(1,2)}$
  &
  16&
  gapped
\\\cline{3-7}%%%%%%%%%%%%%%%
   &
   &
   $A_{0}^{\prime}$
   &
  $\begin{array}{c}\\
        {\rm \widetilde{V}}^{\dagger}\,\,
        {\rm Diag}[1,1,1,0]\,\,
        {\rm V}\\ \\
   \end{array}$
   &
   $    U_{1}^{(1)}\times
        U_{3}^{(1,2)}\times
        U_{1}^{(2)}
   $
   &
   21&
   gapless
\\\cline{3-7}%%%%%%%%%%%%%%%%%
  &
  &
  $A_{1}^{\prime}$
  &
  $\begin{array}{c}\\
        {\rm \widetilde{V}}^{\dagger}\,\,
        {\rm Diag}[1,1,0,0]\,\,
        {\rm V}\\ \\
  \end{array}$
   &
   $    U_{2}^{(1)}\times
        U_{2}^{(1,2)}\times
        U_{2}^{(2)}
   $
   &
  20&
  gapless
\\\cline{3-7}%%%%%%%%%%%%%%%%%%%
  \multirow{-4}{*}[1.25cm]{$\begin{array}{c}
                                n_{1e}=n_{2h}\\\\
                                \epsilon_{Z}=0\\\\\\
                            \end{array}$}&
  \multirow{-4}{*}[1.25cm]{$\begin{array}{c}
                                U_{4}^{(1)}\times
                                U_{4}^{(2)}\\
                            \end{array}$}&
   $A_{2}^{\prime}$
   &
  $\begin{array}{c}\\
        {\rm \widetilde{V}}^{\dagger}\,\,
        {\rm Diag}[1,0,0,0]\,\,
        {\rm V}\\ \\
  \end{array}$
   &
   $    U_{3}^{(1)}\times
        U_{1}^{(1,2)}\times
        U_{3}^{(2)}
   $
    &
  13&
  gapless
\\\cline{3-7}
\hline%%%%%%%%%%%%%%%%%%%%%%%%%%%%%%%%%%%%%%
    $\begin{array}{c}\\
            n_{1e}=n_{2h}\\\\
            \epsilon_{Z}>0\\
     \end{array}$&
    $   U_{2}^{(1\uparrow)}\times
        U_{2}^{(1\downarrow)}\times
        U_{2}^{(2\uparrow)}\times
        U_{2}^{(2\downarrow)}
    $
     &
    $ B^{\prime}$
    &
    $\begin{array}{c}\\
        \left(\begin{array}{cc}
                {\rm 0}&{\rm v}\\
                \widetilde{ {\rm v}}&{\rm 0}\\
            \end{array}\right)\\\\
            {\rm v},
            \widetilde{{\rm v}}\in
            U_{2}\\\\
     \end{array}$
    &
    $       U_{2}^{(1\uparrow,2\downarrow)}\times
            U_{2}^{(1\downarrow,2\uparrow)}
    $
    &
    8&
    gapped
\\
\hline%%%%%%%%%%%%%%%%%%%%%%%%%%%%%%%%%%%%%%%%
    $\begin{array}{c}\\
            n_{1e}>n_{2h}\footnote{Here we assume parameters to be tuned so that a Fermi
surface of electrons only with spin up in layer 1 coincides with the Fermi surface
of holes with spin down in layer 2, for details see Fig. \ref{fig:Apr}.}\\\\
            \epsilon_{Z}> 0\\
     \end{array}$&
    $   U_{2}^{(1\uparrow)}\times
        U_{2}^{(1\downarrow)}\times
        U_{2}^{(2\uparrow)}\times
        U_{2}^{(2\downarrow)}
    $&
    $A_{1}$
    &
    $\begin{array}{c}\\
            \left(\begin{array}{cc}
                    {\rm v}&{\rm 0}\\
                    {\rm 0}&{\rm 0}\\
                  \end{array}\right)\\\\
    {\rm v}\in U_{2}\\\\
         \end{array}$
    &
    $       \,U_{2}^{(1\downarrow)}\times
            U_{2}^{(1\uparrow,2\uparrow)}\times
            U_{2}^{(2\downarrow)}\,
    $
    &
    4&
    gapless
    \\
\end{tabular}
\end{ruledtabular}
\end{table*}

In order to classify phases of the excitonic insulating state we
are going to classify the degeneracy spaces of
order parameters with the same symmetry. For this we consider the
condition (\ref{cond on Delta U1U2H}) and use the method of a singular value decomposition.\cite{linearalgebra}
It allows us to represent an
arbitrary matrix ${\Delta}$ as a product of a unitary matrix
${\rm \widetilde{V}}^{\dagger}$, a diagonal matrix ${\rm D}$
with real non-negative numbers on the diagonal, and another unitary matrix
${\rm V}$. Applying the singular value decomposition to the order parameter at any given values of $s$ and $\mathbf{p}$ we obtain

\begin{equation}
 {\Delta}
 =
 {\rm \widetilde{V}}^{\dagger}_{s}(\mathbf{p})
 {\rm D}_{s}(\mathbf{p})
 {\rm V}_{s}(\mathbf{p}).
 \label{V Diag V}
\end{equation}

However, first of all, we notice that in the considered system, the lowest ground state energy is
realized when matrices ${\rm V},{\rm \widetilde{V}}$ do not
depend on momentum $\mathbf{p}$ and index $s=\pm.$ The reason
for this is that in such a case in the expression for the
ground state energy there is a cancellation  of the product of matrices ${\rm V}$
and ${\rm V}^{\dagger}$ (${\rm \widetilde{V}}$ and
${\rm \widetilde{V}}^{\dagger}$) into a unit matrix.
Such a cancellation leads to the maximal negative contribution to the ground state energy, and therefore the energy of the ground state achieves its minimal value.
If we assume that unitary matrices in Eq. (\ref{V Diag V})
depend on momentum $\mathbf{p}$ and index $s=\pm,$ then unitary matrices at different momenta $\mathbf{p},\mathbf{p}^{\prime}$ and different indices $s,s^{\prime}$ do not cancel each other, which increases the ground state energy comparatively to the previous case.
Thus, we conclude, that in order to realize the lowest energy of the ground state, matrices $\widetilde{{\rm V}}$ and ${\rm V}$ can not depend on the momentum $\mathbf{p}$ and  index $s.$ Therefore the singular value decomposition of the matrix of order parameter becomes

\begin{equation}
 {\Delta}
 =
 {\rm \widetilde{V}}^{\dagger}
 {\rm D}
 {\rm V},
 \label{V Diag V 2}
\end{equation}
where in the right hand side of the equation (\ref{V Diag V 2}) only matrix ${\rm D}$ depends on $\mathbf{p}$ and $s,$ but for brevity we omit these indices.

Second, all transformations from the group $G$, including those
from the factor-space $G/H$, do not change the diagonal elements of
the matrix ${\rm D},$ but change matrices
${\rm \widetilde{V}},$ ${\rm V}$ into any other unitary
matrices. Thus if we introduce the notations
\begin{equation}
 {\rm \widetilde{V}}^{\prime\dagger}
 \equiv
 {\rm U}^{(1)\dagger}
\,
 {\rm \widetilde{V}}^{\dagger},
\qquad
 {\rm V^{\prime}}
 \equiv
 {\rm V}
\,
 {\rm U}^{(2)},
\end{equation}
then under the transformation (\ref{G transformations})-(\ref{cond on Delta1}) the order parameter will transform in the following way:
\begin{equation}
 {\Delta}
 =
 {\rm \widetilde{V}}^{\dagger}
 {\rm D}
 {\rm V}
 ~~\longrightarrow~~
 {\Delta}^{\prime}=
 {\rm \widetilde{V}}^{\prime\dagger}
 {\rm D}
 {\rm V^{\prime}}.
\end{equation}
Thus under this transformation the diagonal matrix ${\rm D}$ does not change. Recall that the degeneracy space of the order parameter is obtained by acting on the order parameter ${\Delta}$ by all transformations from the group $G$ (here transformations from subgroup $H$ will not change the order parameter while remaining transformations from factor-space $G/H$ will create the degeneracy space of the order parameter).
As long as only matrices ${\rm V}$ and ${\rm \widetilde{V}}$ are changed by transformations from $G$, we obtain that the degeneracy space of the order parameter and a phase of the correlated state are determined only by the diagonal elements of the matrix ${\rm D}.$

Finally, from the condition (\ref{cond on Delta U1U2H}) we have found that all
possible degeneracy spaces of the order parameter are classified by numbers of equal and different diagonal elements
in matrix ${\rm D}.$
In the case of physically relevant phases there are additional restrictions on the diagonal elements of matrix ${\rm D}$.
Thus among all possible matrices ${\Delta}$ only physically relevant order parameters satisfy the self-consistency equation.
For the phase classification it is sufficient to consider the BCS self-consistency equation for the order parameter.
Diagonalizing the self-consistency equation by unitary matrices
from Eq. (\ref{V Diag V 2}), one obtains 4 equations for diagonal
elements of the matrix ${\rm D}$, each equation corresponds to
some value of index $\zeta.$
These equations have the same structure and depend on the Fermi momentum $p_{F}$.
If the Fermi momentum $p_{F}$ is the same for all types of electrons (for all indices $\zeta$),
then these 4 self-consistency equations are identical, and apart from a trivial zero solution they have the same
non-zero solution.
Hence in such situation in physically relevant phases the
arbitrary diagonal element in matrix ${\rm D}$ can be equal either to
other non-zero diagonal elements, or be equal to a zero.
The application of the in-plane magnetic field in principle changes such a description because of Zeeman splitting.
However, in the case where the Fermi energy is much greater than the Zeeman energy,
$E_{F}\gg\epsilon_{Z}$, the magnetic field does not change the
situation essentially as long as it is possible to neglect the
difference between Fermi momenta for electrons with opposite spin
projections in the self-consistency equations. Therefore four self-consistency equations on four diagonal elements of the matrix ${\rm D}$ become approximately identical also when relatively small in-plane magnetic field $(\epsilon_{Z}\ll E_{F})$ is applied.

The non-zero solution of these equations is given by the gap function $g_{s}(\mathbf{p})$ which at the Fermi surface $(s=+,|\mathbf{p}|=p_{F})$ determines a gap in the single-particle excitation spectrum. Thus we conclude that in all physically relevant phases the matrix ${\rm D}$ in the singular value decomposition (\ref{V Diag V 2}) of the order parameter $\Delta$ consists of zeros or non-negative diagonal elements which approximately are equal to the gap function $g_{s}(\mathbf{p}).$

Substituting the obtained result into Eq. (\ref{V Diag V 2}), we extract the gap function as a multiplier. Thus we conclude that the order parameter ${\Delta}$ in all physically relevant phases has a form

\begin{equation}
 {\Delta}
  \cong
  g_{s}(\mathbf{p})
 {\rm \widetilde{V}}^{\dagger}
 {\rm D}
 {\rm V}.
  \label{W matrix in Delta}
\end{equation}
Here the matrix ${\rm D}$ is a diagonal matrix with 0 or 1 on the diagonal. The representation (\ref{W matrix in Delta}) becomes approximate in the case of the applied in-plane magnetic field with the condition $\epsilon_{Z}\ll E_{F}$.
The dependence of the order parameter $\Delta$ on variables $s$ and $\mathbf{p}$ is completely given by the function $g_{s}(\mathbf{p}).$

Having the matrix of the order parameter provided,
the symmetry group $H$ is
found from the equation (\ref{cond on Delta U1U2H}). For this the matrix of the order parameter is represented as a single value decomposition (\ref{V Diag V 2}). The constant matrices ${\rm V}$ and ${\rm \widetilde{V}}$
are absorbed into matrices ${\rm U}^{(1)}_{H}$ and ${\rm U}^{(2)}_{H}$ of
the global symmetry transformations from the symmetry group $H.$
Then the equation (\ref{cond on Delta U1U2H}) connects two unitary
matrices ${\rm VU}^{(1)}_{H}{\rm V^{\dagger}}\,$ and
${\widetilde{{\rm V}}{\rm U}}^{(2)}_{H}{\widetilde{{\rm V}}^{\dagger}}$ and the diagonal matrix ${\rm D}$ with $0$ or $1$ on the diagonal.
Thus the matrices ${\rm U}^{(1)}_{H}$ and
${\rm U}^{(2)}_{H}$ of transformations from the symmetry group $H$ are obtained.

\section{\label{sec:5 Phases}Phases}
In this section we provide a detailed description of phases of excitonic insulator state in two-layer graphene system. Results of this section are summarized in Table \ref{tab:phases}.

\subsection{The $B$ phase.}

Firstly we consider the situation when there is no external magnetic field and when the charge carrier densities in layers are the same
$n_{1e}=n_{2h}.$ In such case the symmetry group $G$ of the two-layer Hamiltonian of the system in
normal state is given in Eq. (\ref{G}). Under the mentioned conditions the Fermi circle in the conduction band in layer
1 coincides with the Fermi circle in the valence band in layer 2 due to the electron-hole symmetry in graphene.
Hence the non-vanishing ground state average ${\rm F}$, Eq. (\ref{anom average}), can be formed by all species of electrons. Taking into account that the ground state with the lower energy is more stable, we consider the phase when excitonic correlations are developed among all species of electrons.
In such cases the order parameter matrix $\Delta$ and the matrix ${\rm D},$ Eq. (\ref{W matrix in Delta}),
are not degenerate matrices.
Moreover, because there is only one Fermi circle for all species
of electrons, the most stable ground state is characterized by the matrix ${\rm D}$ in Eq. (\ref{W matrix in Delta}) with
equal non-zero diagonal elements, i.e. ${\rm D}$ is an identity matrix.
As discussed in the previous section, this conclusion follows from the consideration of the self-consistency equation on the order parameter. Thus substituting ${\rm D}= \openone$ into the equation (\ref{W matrix in Delta}) we obtain the following structure for the order parameter in spin$\otimes$valley space:

\begin{equation}
 {\Delta}
 =
 g_{s}(\mathbf{p})
 {\rm V},
  \qquad
 {\rm V}\in U_{4}.
  \label{Delta B phase}
\end{equation}
Such a structure of the order parameter determines the symmetry group $H$ of the ground state and the degeneracy space of the order parameter, and consequently it determines the phase of the excitonic insulator.

\begin{figure}[t!]
\begin{center}
 \includegraphics[width=8.0cm]{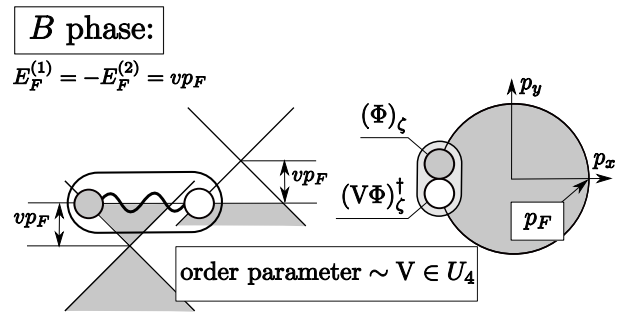}
\end{center}
\caption{
In the $B$ phase within the excitonic paired state
the electron on the Fermi surface in the layer 1 (grey circle) is characterized by the index $\zeta$
in the spin$\otimes$valley basis $\Phi,$ and the absent electron (white circle) in the layer 2 is characterized by the same index $\zeta$ but in the spin$\otimes$valley basis ${\rm V}\Phi,$ which is transformed by the matrix of the order parameter ${\rm V}$.
} \label{fig:Bphase}
\end{figure}

The symmetry group $H$ of the ground state in the considered phase can be found as a group of all unitary transformations ${\rm U}^{(1)}_{H}$, ${\rm U}^{(2)}_{H}$ in layers 1 and 2, which leave the order parameter invariant, Eq. (\ref{cond on Delta U1U2H}).
Solving the condition (\ref{cond on Delta U1U2H}) with the order
parameter (\ref{Delta B phase}) we obtain matrices
${\rm U}^{(1)}_{H}$ and ${\rm U}^{(2)}_{H}$ of symmetry
transformations in the layer 1 and 2 correspondingly.
Thus, having ${\rm U}^{(1)}_{H}$ and ${\rm U}^{(2)}_{H}$,

\begin{equation}
 {\rm U}_{H}^{(1)}={\rm U},\qquad
 {\rm U}_{H}^{(2)}={\rm V}^{\dagger}{\rm UV},
\end{equation}
we can express an arbitrary element of the group $H,$ Eq. (\ref{U1U2 H}),
which transforms electron operators in layers 1 and 2 according to Eq. (\ref{G transformations}). Omitting indices $\zeta,\mathbf{p},s$ we have:
\begin{equation}
    \left(\begin{array}{l}
    a_{1}\\
    a_{2}\\
    \end{array}\right)
    \rightarrow
 \left(\begin{array}{cc}
 {\rm U}&{\rm 0}\\
 {\rm 0}&{\rm V^{\dagger}UV}\\
 \end{array}\right)
    \left(\begin{array}{l}
    a_{1}\\
    a_{2}\\
    \end{array}\right).
    \label{transform of a}
\end{equation}
Here the matrix ${\rm V}$ is given by the fixed matrix of the order parameter (\ref{Delta B phase}).
The unitary matrix ${\rm U}$ is present in transformations in both layers. This means that the symmetry group $H$ of the ground state in the considered phase (\ref{Delta B phase}) consists of the combined transformations in both layers 1 and 2. The unitary group of the combined transformations in layers 1 and 2 is denoted as $U^{(1,2)}_{4},$

\begin{equation}
  \left(\begin{array}{cc}
 {\rm U}&{\rm 0}\\
 {\rm 0}&{\rm V^{\dagger}UV}\\
 \end{array}\right)
 \in
 U^{(1,2)}_{4}\equiv H.
\label{transform of a element H}
\end{equation}

The combined transformations from the group $U_{4}^{(1,2)}$ can also be described in terms of generators of these transformations. For this, each element of the group is written as the
exponential function of the element of the group's algebra

\begin{equation}
 \left(\begin{array}{cc}
 {\rm U}&{\rm 0}\\
 {\rm 0}&{\rm V^{\dagger}UV}\\
 \end{array}\right)
 =
 \exp\left[i\vec{\theta}\vec{\Gamma}_{H}\right]
 \in H,
 \label{def of generators}
\end{equation}
where
$\vec{\theta}$ is a vector of real variables and a vector
$\vec{\Gamma}_{H}$ consists of generators of the group $H.$ For the considered phase these
generators are:

\begin{equation}
 \left(\Gamma_{H}\right)_{m}\equiv
 \left(\begin{array}{cc}
        \lambda_{m}&0\\
        0&{\rm V^{\dagger}}\lambda_{m}{\rm V}\\
        \end{array}
  \right),\quad
  m=0,1,...,15.
  \label{B phase generators}
\end{equation}
In contrast to Eq.(\ref{B phase generators}) transformations, which change the order parameter and create a degeneracy space $G/H,$ are described by the following generators:

\begin{equation}
 \left(\Gamma_{G/H}\right)_{m}\equiv
 \left(\begin{array}{cc}
        \lambda_{m}&0\\
        0&-{\rm V^{\dagger}}\lambda_{m}{\rm V}\\
        \end{array}
  \right),\quad
  m=0,1,...,15.
  \label{B phase generators G/H}
\end{equation}
Here the $2\times2$ block matrices $\left(\Gamma_{H}\right)_{m}$ and $\left(\Gamma_{G/H}\right)_{m}$ act in the space of layers 1 and 2, the matrix $\lambda_{m}$ acts on the spin$\otimes$valley basis in the layer 1 and matrices $\pm{\rm V^{\dagger}}\lambda_{m}{\rm V}$ acts on the basis $\Phi$ in layer 2. The spin$\otimes$valley basis $\Phi$ is the same in both layers. Matrices $\lambda_{m}$ are 4$\times$4 Hermitian traceless matrices of generators of transformations from the unitary group $U_{4}.$
The total number of generators $\Gamma_{G/H},$ Eq. (\ref{B phase generators G/H}), equals to the dimension of the degeneracy space $G/H,$ ${\rm dim}[G/H]$. In the $B$ phase ${\rm dim}[G/H]={\rm dim}[G]- {\rm dim}[H]=32-16=16.$

The electron operators in the second layer can be transformed by the matrix of the order parameter: ${\rm V}a_{2}\rightarrow a_{2}^{\prime}$, see Eqs. (\ref{Delta B phase}) and (\ref{mean-field H}), or, equivalently, the spin$\otimes$valley basis in layer 2 can be transformed by the matrix of the order parameter. In such a case from Eqs. (\ref{transform of a}) and (\ref{B phase generators}) it follows, that the transformation from the group $H,$ in contrast to the transformation from $G/H,$ can be represented by identical transformations in both layers.
These identical transformations act
by the same matrix ${\rm U}$ on the spin$\otimes$valley basis
$\Phi$ in the layer 1 and on the transformed spin$\otimes$valley basis
${\rm V}\Phi$ in the layer 2, Fig. \ref{fig:Bphase}.
Hence the matrix of the order parameter (\ref{Delta B phase}) defines
the relative unitary rotation of the spin$\otimes$valley basis
$\Phi$ in layer 2 with respect to layer 1.
It signifies the relative symmetry breaking: the ground state is not invariant under unitary transformations of the basis $\Phi$ in one layer relatively to the basis $\Phi$ in another layer. The basis $\Phi$ in layer 1 is "locked" relatively to the basis $\Phi$ in layer 2 by the matrix of the order parameter which defines the relative unitary rotation of one basis with respect to another.

Because of the presence of the relative symmetry breaking by the
order parameter (\ref{Delta B phase}) the phase discussed here resembles the superfluid $B$ phase in the liquid Helium-3.\cite{mineevsamokhin,voloviksymmetryin3-Hechapter,leggettrmp75,wheatley,mineevufn}

In $B$ phase the matrix of the order parameter is not degenerate. It means that all species of charge carriers develop excitonic correlations, therefore a single particle excitation spectrum is gapped.

The external conditions ($\epsilon_{Z}=0,$ $n_{1e}=n_{2h}$) for the $B$ phase can be violated by an in-plane magnetic field or by external gates. However the excitonic correlations continue to exist in the $B$ phase until the difference of radiuses of Fermi circles is bigger than $2g_{+}(p_{F})/v,$ where $g_{+}(p_{F})$ is a gap in a single-particle excitation spectrum in the $B$ phase.
Indeed such behavior can be seen, if one creates an asymmetry between charge carriers densities in layers, which can be expressed in terms of a shift $\delta E_{F}>0$ of Fermi enetgies: $E_{F}^{(1)}=E_{F}+\delta E_{F},$ $E_{F}^{(2)}=-E_{F}+\delta E_{F}$. Substituting these values to the mean field Hamiltonian (\ref{mean-field H}) and finding its eigenvalues, one obtains\cite{mineevsamokhin} two branches of excitation spectrum
$
\varepsilon_{s}^{(\pm)}(p)=\sqrt{(svp-E_{F})^2+g_{s}^{2}(p)}\pm \delta E_{F}.
$
At values $\delta E_{F}=g_{+}(p_{F})$ one of branches of excitation spectrum becomes zero at $s=+,p=p_{F}.$ At this situation the excitonic pairing stops being energetically favorable and the system appears in the normal state via a first order phase transition. In similar way, when Fermi circles for charge carriers with opposite spin projection are separated by $2g_{+}(p_{F})/v$ due to an in-plane magnetic field, excitonic correlations between charge carriers on these Fermi circles vanish.
This fact is schematically shown in the phase diagram Fig.\ref{fig 1}b: at the borders of the $B$ phase in the phase diagram excitonic correlations are no longer energetically stable and the excitonic insulator state transforms into either a normal state or into another phase via a first order phase transition.

\subsection{The $A_{0}^{\prime},A_{1}^{\prime},A_{2}^{\prime}$ phases.}
In this section we consider phases under the same external conditions as in $B$ phase, thus the symmetry group $G$ is again given by Eq.(\ref{G}). We consider phases where order parameters are characterized by the degenerate matrices of rank $r<4.$ In such phases only a part of electron species develop excitonic correlations, therefore the single particle excitation spectrum is gapless for certain species of  electrons. The matrix of the order parameter can be chosen as follows, compare with Eq.(\ref{W matrix in Delta}):
\begin{equation}
 {\Delta}
 =
 g_{s}(\mathbf{p})
 \widetilde{{\rm V}}^{\dagger}
 {\rm Diag}[a,b,c,0]
 {\rm V},
  \qquad
 \widetilde{{\rm V}}^{\dagger},{\rm V}\in U_{4}.
  \label{Delta A1pr phase}
\end{equation}
Here the diagonal matrix ${\rm Diag}$ determines the order parameter in phases, which are denoted as $A_{0}^{\prime}$, $A_{1}^{\prime}$, $A_{2}^{\prime}$: numbers $(a,b,c)$ are given by $(1,1,1)$ in $A_{0}^{\prime}$ phase, $(1,1,0)$ in $A_{1}^{\prime}$ phase and $(1,0,0)$ in $A_{2}^{\prime}$ phase. Using the transformed electron operators ${\rm\widetilde{V}}a_{1}$ in the layer 1 and ${\rm V}a_{2}$ in the layer 2, see Eqs. (\ref{mean-field H}) and (\ref{Delta A1pr phase}), the self-consistency equation on the order parameter becomes diagonal, and only first $r$ out of four equations for diagonal elements will have non-zero solutions. We assume that in the self-consistency equations we can use the screened interaction among charge carriers in the system in normal state.\cite{KharitonovEfetov0903} In this case self-consistency equations in $A_{0}^{\prime},A_{1}^{\prime},A_{2}^{\prime}$ phases are identical to self-consistency equations in $B$ phase, therefore their non-zero solutions are given by the same gap function $g_{s}(\mathbf{p}).$

Substituting the order parameter in the symmetry
condition (\ref{cond on Delta U1U2H}) one obtains matrices of
symmetry transformations in layer 1 and 2, for example for $A_{1}^{\prime}$ phase one gets

\begin{equation}
 {\rm U}^{(1)}_{H}=
        {\rm \widetilde{V}}^{\dagger}
        \left(\begin{array}{cc}
                {\rm u}&0\\
                0&{\rm u}^{\prime}\\
                \end{array}
        \right){\rm \widetilde{V}},
 \quad
 {\rm U}^{(2)}_{H}=
        {\rm V}^{\dagger}
        \left(\begin{array}{cc}
                \rm {\rm u}&0\\
                0&{\rm u}^{\prime\prime}\\
                \end{array}
        \right){\rm V},
        \label{U1U2 A1phase}
\end{equation}
where
\begin{equation}
 {\rm u}\in U_{2}^{(1,2)},\quad
 {\rm u}^{\prime}\in U_{2}^{(1)},\quad
 {\rm u}^{\prime\prime}\in U_{2}^{(2)}.
        \label{U1U2 A1phase2}
\end{equation}
Here, similarly to the $B$ phase, the $2\times2$ unitary matrix ${\rm u}$ determines
the combined unitary rotations of the first two components of the
spin$\otimes$valley basis ${\rm \widetilde{V}}\Phi$ in layer 1 and the
first two components of the spin$\otimes$valley basis ${\rm V}\Phi$ in
layer 2. Therefore such phase is characterized by a partial relative
symmetry breaking. Remaining matrices
${\rm u}^{\prime},{\rm u}^{\prime\prime}\in U_{2}$ determine
independent unitary rotations of the other 2 components of
corresponding spin$\otimes$valley basis in layers. These other 2
components correspond to quasiparticle's states which do not
contribute to the excitonic condensation, their single particle excitation spectrum is gappless. Therefore only 2 out of 4 electron's species are involved in the excitonic condensation. Because of
this, such a phase is similar to the superfluid $A_{1}$ phase of
liquid Helium-3, where paired states with only one spin projection
$S_{z}=+1$ are present in the condensate.\cite{mineevsamokhin}
The $A_{1}$ phase of Helium-3 exists only in magnetic field, in
order to underline the stability of the phase in the absence of the magnetic field
we denote the phase discussed here
by an additional prime, therefore it is denoted as the $A_{1}^{\prime}$ phase of the excitonic
insulator. Other phases with degenerate matrices of order
parameters are denoted as $A_{0}^{\prime}$ and $A_{2}^{\prime}.$

In phases $A_{0}^{\prime},$ $A_{1}^{\prime},$ $A_{2}^{\prime}$ the number of non-zero diagonal elements in the diagonal matrix ${\rm Diag}$ determines the rank $r$ of the symmetry group of combined unitary rotations, denoted as $U_{r}^{(1,2)}.$ Zeros in the diagonal of the matrix ${\rm Diag}$ correspond to electron states which do not develop excitonic correlations, and, therefore, these states can be unitary transformed independently in each layer.

Consequently the symmetry group $H$ for $A_{0}^{\prime},$ $A_{1}^{\prime},$ $A_{2}^{\prime}$ phases can be easily identified. For example, the symmetry group $H$ for $A_{1}^{\prime}$ phase is following:
\begin{equation}
 H=U_{2}^{(1)}\times
 U_{2}^{(1,2)}\times
 U_{2}^{(2)}.
\end{equation}
The dimension of the degeneracy space is calculated as follows:
for $A_{0}^{\prime}$ phase ${\rm dim}[G/H]=32-1-9-1=21;$
for $A_{1}^{\prime}$ phase ${\rm dim}[G/H]=32-3\times4=20;$
for $A_{2}^{\prime}$ phase ${\rm dim}[G/H]=32-9-1-9=13.$

\subsection{The $B^{\prime}$ phase.}
In this section we consider the two-layer graphene system in
an in-plane magnetic field.
Our analysis is based on the comprehensive study by Aleiner and co-authors\cite{Aleiner} of the spontaneous symmetry breaking in graphene subjected to an in-plane magnetic field. When an in-plane magnetic field is applied, the
Fermi circles for quasiparticles with different spin projections
become separated due to a Zeeman splitting.
Such splitting changes the symmetry group $G,$ Eq. (\ref{G}), of the initial Hamiltonian $\hat{H}_{{\rm 2layer}}$ toward a direct product of
4 unitary groups $U_{2},$

\begin{equation*}
G=U_{2}^{(1\uparrow)}\times
        U_{2}^{(1\downarrow)}\times
        U_{2}^{(2\uparrow)}\times
        U_{2}^{(2\downarrow)}.
\end{equation*}
Each of these $U_{2}$ groups transforms a valley space of electrons with corresponding spin projections in one layer, e.g. $U_{2}^{(1\uparrow)}$ transforms electrons with spin up in layer 1.

\begin{figure}[t]
\begin{center}
 \includegraphics[width=8.0cm]{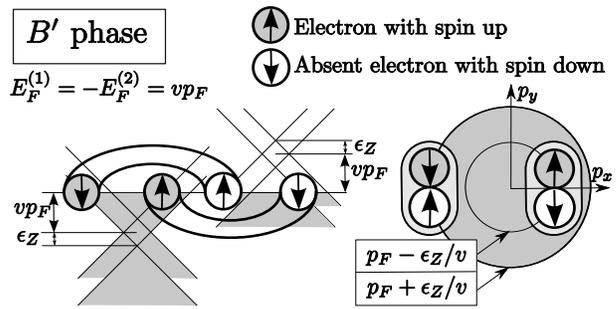}%width=\columnwidth
 \caption{
Excitonic correlations in the two-layer graphene system with an in-plane
magnetic field $h$ in the case of equal charge densities in layers
$n_{1e}=n_{2h}.$ Because of a Zeeman splitting $2\epsilon_{Z}$ there are two Fermi
circles with radiuses $p_{F}\pm\epsilon_{Z}/v.$ The absent electron with a particular
spin projection is considered as a quasiparticle (hole) with an opposite spin
projection.}
 \label{fig:Bpr phase}
\end{center}
\end{figure}

The $B^{\prime}$ phase can be obtained from the $B$ phase by the application of an in-plane magnetic field. Such magnetic field should be big enough to break the excitonic correlations in the $B$ phase and to split Fermi circles.
Therefore a Zeeman energy $\epsilon_{Z}$
should be bigger than a gap in the excitation spectrum in the
$B$ phase, $\epsilon_{Z}>g_{+}(p_{F}).$ In such a case, due to the initial equality of charge carrier densities
$n_{1e}=n_{2h}$ in the $B$ phase, the two Fermi circles in layer 1 coincide with two
Fermi circles in layer 2. Thus it leads to the appearance of two
different Fermi circles in the system, Fig. \ref{fig:Bpr phase}.
Consequently, the electron-hole pairs, which appear on different
Fermi circles, have different properties: thus such electron-hole pairs have different spin projection, $+1$ or $-1,$ Fig. \ref{fig:Bpr phase}.
Also due to slightly different Fermi momenta, electron-hole pairs on different Fermi circles are
characterized by slightly different gap functions.
Hence in the spin$\otimes$valley basis $\Phi,$

\begin{equation}
 \Phi=
 \left(
 \uparrow \mathbf{K}_{+},
 \uparrow \mathbf{K}_{-},
 \downarrow \mathbf{K}_{+},
 \downarrow \mathbf{K}_{-}
 \right),
 \label{Phi}
\end{equation}
the order parameter has the following structure (compare with Eq. (\ref{Delta B phase}))

\begin{equation}
 {\Delta}=
 \left(\begin{array}{cc}
    {\rm 0}&g^{\prime}_{s}(\mathbf{p}){\rm v}\\
    g^{\prime\prime}_{s}(\mathbf{p}){\rm \widetilde{v}}&{\rm 0}\\
    \end{array}\right).
    \label{Delta Bpr exact}
\end{equation}
Here matrices ${\rm v}$ and ${\rm \widetilde{v}}$ are
unitary $2\times2$ matrices, which, by analogy with the $B$ phase, determine the relative unitary
rotation of a valley space of electron states with one spin
projection in layer 2 relatively to electron states with another spin
projection in layer 1.
Functions $g^{\prime}_{s}(\mathbf{p}), g^{\prime\prime}_{s}(\mathbf{p})$ are gap functions, which differ from
each other only because of the presence of a Zeeman splitting.
However when the Fermi energy is much bigger than Zeeman energy,
$E_{F}\gg\epsilon_{Z},$ the difference between these functions is
negligible and they are approximately equal to the gap function in the $B$ phase, $g^{\prime}_{s}(\mathbf{p})\approx g^{\prime\prime}_{s}(\mathbf{p})\approx g_{s}(\mathbf{p}).$ Thus,

\begin{equation*}
 \Delta\approx g_{s}({\mathbf p})
  \left(\begin{array}{cc}
    {\rm 0}&{\rm v}\\
    {\rm \widetilde{v}}&{\rm 0}\\
    \end{array}\right).
\end{equation*}
Using this approximation, the symmetry group $H$ of the order parameter can be found from the
condition (\ref{cond on Delta U1U2H}). As a result one obtains matrices ${\rm U}^{(1)}_{H}$ and ${\rm U}^{(2)}_{H}$ of
the transformations (\ref{U1U2 H}) from the group $H$ in the layer 1 and 2
respectively (matrices are written in the basis (\ref{Phi}) in each layer):

\begin{equation}
 {\rm U}^{(1)}_{H}=
    \left(\begin{array}{cc}
          {\rm u}&0\\
          0&\widetilde{{\rm u}}\\
    \end{array}\right),
 \quad
 {\rm U}^{(2)}_{H}=
    \left(\begin{array}{cc}
          {\rm \widetilde{v}}^{\dagger}{\rm \widetilde{u}\widetilde{v}}&0\\
          0&{\rm v}^{\dagger}{\rm uv}\\
    \end{array}\right).
    \label{U1U2 Bpr phase}
\end{equation}
The corresponding electron operators are transformed as follows:
\begin{equation}
 a_{1,\uparrow}\rightarrow{\rm u}a_{1,\uparrow},\quad
 a_{2,\downarrow}\rightarrow{\rm v}^{\dagger}{\rm uv}a_{2,\downarrow},\quad
 {\rm u}\in U_{2}^{(1\uparrow,2\downarrow)},
 \label{Bpr u}
\end{equation}
\begin{equation}
 a_{1,\downarrow}\rightarrow{\rm \widetilde{u}}a_{1,\downarrow},\quad
 a_{2,\uparrow}\rightarrow{\rm \widetilde{v}}^{\dagger}{\rm \widetilde{u}\widetilde{v}}
 a_{2,\uparrow},\quad
 {\rm \widetilde{u}}\in U_{2}^{(1\downarrow,2\uparrow)}.
 \label{Bpr u tilde}
\end{equation}
The unitary $2\times2$ matrix ${\rm u},$ Eq. (\ref{Bpr u}), determines a subgroup
of the group $H,$ which consists of combined unitary rotation of
valley space of electrons with spin up in layer 1 and electrons
with spin down in layer 2: ${\rm u}\in
U_{2}^{(1\uparrow,2\downarrow)}.$ The unitary $2\times2$ matrix
${\rm \widetilde{u}},$ Eq. (\ref{Bpr u tilde}), defines another corresponding subgroup
of the group $H,$ ${\rm \widetilde{u}}\in
U_{2}^{(1\downarrow,2\uparrow)}\subset H.$
Hence in the phase considered here the symmetry group $H$ of the order parameter is given by direct product of two subgroups
\begin{equation}
 H=U_{2}^{(1\uparrow,2\downarrow)}\times
U_{2}^{(1\downarrow,2\uparrow)}.
\end{equation}
Using expressions for groups $G$ and $H$ in the $B^{\prime}-$phase, we found that the degeneracy space $G/H$ is
8-dimensional, ${\rm dim}[G/H]=4\times4-4-4=8.$ It is also defined by the structure of the order
parameter (\ref{Delta Bpr exact}), i.e. here the degeneracy space is determined as a space of all possible unitary $2\times2$ matrices
${\rm v}$ and ${\rm \widetilde{v}}.$ Because of the non-degenerate matrix of the order parameter, the single particle excitation spectrum in this phase is gapped.

Similarly to the $B$ phase, the excitonic correlations in the $B^{\prime}$ phase cease to exist when the external conditions ($\epsilon_{Z}>g_{+}(p_{F}),$ $n_{1e}=n_{2h}$) are perturbed, i.e. when Fermi circles in different layers are separated for the energy interval which is bigger than a double value of a gap in the single-particle excitation spectrum.
Thus, in particular, in the schematic phase diagram Fig. \ref{fig 1}(b) at the border of the $B^{\prime}-$phase (when a symmetry $n_{1e}=n_{2h}$ of charge carriers densities is violated) the ground state of the system transforms to an uncorrelated normal ground state via the first order phase transition.

\subsection{The $A_{1}$ phase.\label{sec:5 A1 phase}}
In contrast to $B$ and $B^{\prime}$ phases, where all species of charge carriers develop excitonic correlations, in this subsection we discuss another possible realization of the excitonic insulator state in the two-layer graphene system. We show that the excitonic correlated state can exist in the presence of an in-plane magnetic field and a specially chosen asymmetry in charge carriers densities in layers.

\begin{figure}[t]
\begin{center}
 \includegraphics[width=8.0cm]{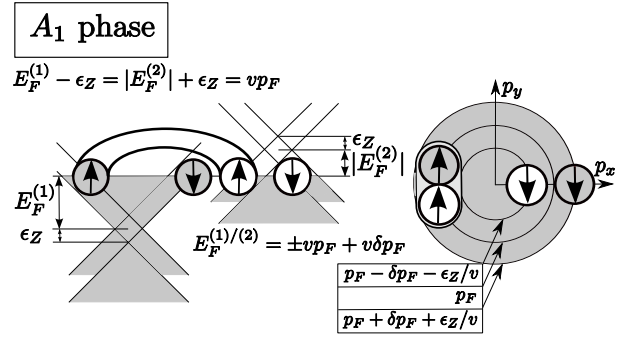}%{ex3figapr.eps}%width=\columnwidth
 \caption{
Excitonic correlations in the $A_{1}$ phase.
Starting from a zero magnetic field the asymmetry between charge carrier densities $n_{1e}>n_{2h}$ is created. In terms of Fermi energies it means $E_{F}^{(1)}>|E_{F}^{(2)}|.$
A magnitude of an in-plane magnetic field is chosen such that a Zeeman energy $\epsilon_{Z}$ satisfies the condition: $E_{F}^{(1)}-\epsilon_{Z}=|E_{F}^{(2)}|+\epsilon_{Z},$ where $(E_{F}^{(1)}-\epsilon_{Z})/v$ is the radius of the Fermi circle for electrons with spin up in layer 1, and $(|E_{F}^{(2)}|+\epsilon_{Z})/v$ is a radius of the Fermi circle for electrons with spin up in layer 2. Both of these Fermi circles are situated at the same Fermi momentum $p_{F}$.
Therefore two out of four Fermi circles coincide, leading to excitonic
correlations between only half of electron's species.
Using the expression for the Fermi energies $E_{F}^{(1)/(2)}=\pm vp_{F}+v\delta p_{F},$ $p_{F}>\delta p_{F},$ where $\delta p_{F}=(\sqrt{n_{1e}}-\sqrt{n_{2h}})\sqrt{\pi}/2,$ the condition on the Zeeman energy is the following:
$\epsilon_{Z}=v\delta p_{F}.$
 }
 \label{fig:Apr}
\end{center}
\end{figure}

In order to achieve a necessary external conditions, firstly we consider the two-layer system without a magnetic field and with equal charge carrier densities in layers. Under such conditions the spectrum of electrons in both layers have only one Fermi circle at the Fermi momentum $p_{F}.$ By changing the external gate voltages we create asymmetry between charge carrier densities in layers: $n_{1e}>n_{2h}.$
Thus the Fermi circle in the layer 1 is situated at the momentum $p_{F}+\delta p_{F}$,
and the Fermi circle in the layer 2 is situated at the momentum $p_{F}-\delta p_{F},$ where $\delta p_{F}>0,$ and $p_{F}$ is the Fermi momentum in the case $n_{1e}=n_{2h}$ (i.e. in $B$ and $B^{\prime}$ phases).
It is assumed that the separation between Fermi circles is big enough to prevent the development of excitonic correlations.
Keeping the chosen values of densities, we switch-on an in-plane magnetic field with such a magnitude that the Zeeman energy $\epsilon_{Z}$ is equal to the energy shift of each Fermi surface, $\epsilon_{Z}=v\delta p_{F},$ Fig. \ref{fig:Apr}. The presence of an in-plane magnetic field signifies that the symmetry group $G$ in such case is the same as in $B^{\prime}$ phase.

The external conditions mentioned above lead to the situation when only two out of four Fermi circles coincide: both the Fermi circle of electrons with spin up in the layer 1 (a Fermi circle with a smaller radius in the layer 1) and the Fermi circle of electrons with spin up in the layer 2 (a Fermi circle with a bigger radius in the layer 2) are situated at the same Fermi momenta $p_{F}.$
Thus electron-hole pairs are formed on these two coincided Fermi circles.
Notice, that a total spin projection of such an electron-hole pair is equal to a zero in contrast to electron-hole pairs with spin projections $+1$ or $-1$ in the $B^{\prime}$ phase.
Fermi circles for electrons with spin down in both layers do not coincide with any other Fermi surfaces. Therefore corresponding electrons and holes are in a normal state (i.e. they do not participate in excitonic correlations), their single-particle excitation spectrum is gapless.

Thus for the phase discussed here only half of all electron species in the system develop excitonic correlations. It is reflected in the order parameter, whose structure in the basis (\ref{Phi}) in both layers is given by the following expression:

\begin{equation}
 \Delta=
 g_{s}(\mathbf{p})
 \left(\begin{array}{cc}
 {\rm v}&0\\
 0&0\\
 \end{array}\right).
 \label{Delta A1}
\end{equation}
Here the gap function $g_{s}({\mathbf{p}})$ is the same as in other phases due to the same Fermi momentum $p_{F}$ in the self-consistency equation and due to the approximation of the interaction among charge carriers in all phases by the screened interaction among charge carriers in the system in normal state. Similarly to the phases discussed previously, the unitary $2\times2$ matrix ${\rm v}$ in the order parameter (\ref{Delta A1}) determines a relative unitary rotation of the valley space of electrons with spin up in layer 2 relatively electrons with spin up in layer 1.

Solving the condition (\ref{cond on Delta U1U2H}) with the order parameter (\ref{Delta A1}) we found that transformations from the group $H$ are represented in layer 1 and 2 by following matrices ${\rm U}^{(1)}_{H}$ and ${\rm U}^{(2)}_{H}$ respectively (both matrices are written in the basis (\ref{Phi}) in each layer):

\begin{equation}
 {\rm U}^{(1)}_{H}=
    \left(\begin{array}{cc}
          {\rm u}&0\\
          0&{\rm u}^{\prime}\\
          \end{array}
    \right),
    \quad
 {\rm U}^{(2)}_{H}=
    \left(\begin{array}{cc}
          {\rm v^{\dagger}uv}&0\\
          0&{\rm u}^{\prime\prime}\\
          \end{array}
    \right).
    \label{U1U2 A1pr phase}
\end{equation}
Here matrices ${\rm u},{\rm u}^{\prime},{\rm u}^{\prime\prime}$ are unitary $2\times2$ matrices.
The matrix ${\rm u}$ performs a combined unitary transformation of a valley space of electrons with spin up in both layers, in addition the valley space of electrons in layer 2 are rotated by the order parameter (\ref{Delta A1}), compare with Eq. (\ref{Bpr u}). The valley space of electrons with spin down is transformed by the unitary matrix ${\rm u}^{\prime}$ in layer 1 and by the unitary matrix ${\rm u}^{\prime\prime}$ in layer 2 correspondingly:

\begin{align}
 a_{1,\uparrow}\rightarrow{\rm u}a_{1,\uparrow},\quad
 &a_{2,\uparrow}\rightarrow{\rm v}^{\dagger}{\rm uv}a_{2,\uparrow},\quad
 &&{\rm u}\in U_{2}^{(1\uparrow,2\uparrow)},
 \label{A1 u}
\\\nonumber\\
 &a_{1,\downarrow}\rightarrow{\rm u^{\prime}}a_{1,\downarrow},\quad
 &&{\rm u^{\prime}}\in U_{2}^{(1\downarrow)},
  \label{A1 ut}
\\\nonumber\\
 &a_{2,\downarrow}\rightarrow{\rm u^{\prime\prime}}a_{2,\downarrow},\quad
 &&{\rm u^{\prime\prime}}\in U_{2}^{(2\downarrow)}.
 \label{A1 utt}
\end{align}
Thus the group $\textsl{H}$ consists of the direct product of 3
unitary groups,
\begin{equation}
 H=U_{2}^{(1\downarrow)}\times
 U_{2}^{(1\uparrow,2\uparrow)}\times
 U_{2}^{(2\downarrow)}.
 \label{A1 H}
\end{equation}
Using the expressions for initial symmetry group $G$ and Eq. (\ref{A1 H}), we find that the degeneracy space $G/H$ in this phase is 4 dimensional, ${\rm dim}[G/H]=16-3\times4=4.$. It is determined by the manifold of all possible
matrices ${\rm v}\in U_2$ in the structure of the order parameter (\ref{Delta A1}).

The subgroups $U_{2}^{(1\downarrow)}$ and $U_{2}^{(2\downarrow)}$ are present in both groups $G$ and $H,$ the appearance of the correlated state does not change them.
Therefore for the phase discussed here the initial symmetry
is broken only partially.
According to the
similarities with the superfluid $A_{1}$ phase of liquid Helium-3 (i.e. that the phase described
here exists in magnetic field and has a partial relative symmetry
breaking\cite{mineevsamokhin,voloviksymmetryin3-Hechapter,leggettrmp75,wheatley,mineevufn}), the phase discussed in this subsection was denoted as $A_{1}$ phase.

\section{\label{sec:6 Results}Results and Discussions}

In the present paper we consider a two-layer graphene system where external gate voltage induces a finite density of electrons in one layer and holes in another. Assuming that the transition temperature $T_{c}$ towards excitonic insulator is high enough so that it can be observed, we classify phases of such correlated state.
In order to obtain different excitonic correlations and therefore different phases we propose to use parallel to graphene layers magnetic field and perpendicular to graphene layers electric field.

Firstly we consider the Hamiltonian of the two-layer graphene system. We recognize that the ground state is characterized by a high symmetry group - the group of unitary rotations of spin$\otimes$valley space of electrons in each layer independently.
Below a transition temperature $T_{c}$ this symmetry is reduced by a non-zero order parameter towards a symmetry of the excitonic insulating ground state, which consists of electron-hole pairs with electrons on one layer and holes on another.
Following the BCS theory of superconductivity, we identify the condition for such electron-hole pairing, determine the order parameter and build a BCS-like mean-field theory of the excitonic insulator.
Analyzing a symmetry breaking of the initial ground state by the order parameter, we consider a condition that mutually determines the order parameter and the corresponding symmetry group of the excitonic insulator ground state.
Using a singular value decomposition of the matrix of the order parameter, for each phase of the excitonic insulator we obtain a corresponding symmetry group of the ground state, a structure of the order parameter and its degeneracy space.
The results of a phase classification of the excitonic insulator are shown in Table \ref{tab:phases}, the most energetically stable phases are shown in the phase diagram, Fig. \ref{fig 1}(b), and on Figs. \ref{fig:Bphase}, \ref{fig:Bpr phase}, \ref{fig:Apr}.

It is important to notice that the excitonic correlations in all phases discussed in this paper origin from the coincided Fermi surfaces at approximately the same Fermi momentum $p_{F}$ (we use assumption $E_{F}\gg \epsilon_{Z}$, where the Fermi energy $E_{F}$ is determined in the system without a magnetic field and with equal densities of charge carriers in layers $n_{1e}=n_{2h}$). Thus assuming that the interaction among charge carriers is the same in all phases (i.e. that the effect of excitonic correlations on the screening of the interaction can be neglected\cite{KharitonovEfetov0903}), we obtain that the energy gap in the single particle excitation spectrum in all phases is determined by the same self-consistency equation. Therefore the transition temperature $T_c$ estimated from the self-consistency equation\cite{mineevsamokhin} should be the same for all phases.

At a temperature lower than the transition temperature $T_{c}$, transitions between phases in the phase diagram
are found to be of the first order. Phases of excitonic insulator have different properties: thus the electron-hole pairs in $B^{\prime}$ phase have total spin projection $+1$ or $-1,$ Fig. \ref{fig:Bpr phase}, whereas in the $A_{1}$-phase a total spin projection of an electron-hole pair is equal to zero, Fig. \ref{fig:Apr}.
%We show, that the excitonic insulator ground state can have stable topological defects - vortices. Their properties and realization on experiment are discussed in Refs. \cite{seradjeh weber franz,shevchenko 1997}.

According to the number of
estimations of the critical temperature in the considered
system, the most optimistic estimation gives values of $T_{c}$
close to a room temperature.\cite{min .. macdonald} However this
estimation\cite{min .. macdonald} does not take screening of the
Coulomb interaction into account, explaining it by the assumption
of the first order phase transition in the system. Some other
estimations\cite{kharitonov efetov,KharitonovEfetov0903} point on the improbability of observation of excitonic
condensation due to extremely low transition temperature $\lesssim
1{\rm mK},$ $(T_{c}\approx 10^{-7}E_{F})$. According to Refs. \cite{kharitonov efetov,KharitonovEfetov0903} the reason
for low transition temperature lays in the effective screening of
the Coulomb interaction by a big number $N$ of species of electrons.\cite{kharitonov efetov,kharitonov efetov,foster aleiner} In the
considered system  $N=8$, which is given by product of 2 valleys,
2 spin projections and 2 layers. Such large $N$ increases screening
and makes excitonic condensation not so effective, as in the
monolayer graphene \cite{khveshchenko} and especially in
the monolayer graphene in a magnetic field,\cite{gorbar,Aleiner}
where $T_{c}$ can reach value up to $10^{-4}~E_{F}.$ However,
recent investigations, based on a detailed treatment of the screened
Coulomb interaction \cite{lozovik2009,lozovik2009v2,lozovik2010,lozovik2012,sodemann macdonald} and on a consideration of a multi-band pairing
\cite{mink,lozovik2012} or pairing with nonzero momentum\cite{efimkin nonzer mom} suggest that the transition temperature $T_{c}$ can be sufficiently big for the experimental observation of the excitonic insulator in the considered system.
Together with recent experimental realization of the two-layer graphene system \cite{schmidt,schmidt2,schmidt3,schmidt4,FalkoGaugeField,FalkoCheianovTunable} it
provides a hope that the phase diagram of the excitonic insulator, Fig. \ref{fig 1}(b), under favorable conditions\cite{su macdonald,bistritzer macdonald,basu,efimkin disoder} will be observed experimentally.

\begin{acknowledgments}
Y.F.S. would like to thank L. Glazman, E. Burovski and Y. Sherkunov for
useful discussions. The authors thank EPSRC and Physics Department at Lancaster University for financial support.
\end{acknowledgments}

\end{document}